\documentclass[aps,prc,showpacs,floatfix,footinbib,twocolumn,superscriptaddress]{revtex4-1}

\usepackage{amsmath}
\usepackage{stackengine}
\usepackage[dvips]{graphics,graphicx}
\usepackage{graphicx}% Include figure files
\usepackage{dcolumn}% Align table columns on decimal point
\usepackage{bm}% bold math

\usepackage{xcolor}
\begin{document}

\preprint{APS/123-QED}

\title{Normalised symmetric cumulants as a measure of QCD phase transition: a viscous hydrodynamic study}%  ARV-HD} Force line breaks with \\

\author{Ashutosh Dash}
 \email{ashutosh.dash@niser.ac.in}
\author{Victor Roy}%
 \email{victor@niser.ac.in}
\affiliation{%
National Institute of Science Education and Research,
HBNI, 752050 Odisha, India.\\
}%

\begin{abstract}
Finding the existence and the location of the QCD critical point is one of the main goals of the RHIC beam energy scan program. 
To make theoretical predictions and corroborate with the experimental data requires modeling the space-time evolution of the matter created
in heavy-ion collisions by dynamical models such as the relativistic hydrodynamics with an appropriate Equation of State (EoS). 
In the present exploratory study, we use a viscous 2+1 dimensional event-by-event (e-by-e) hydrodynamic code 
at finite baryon densities with two different EoSs (i) Lattice QCD + HRG with a crossover transition and (ii) EoS with a first-order phase transition to studying the normalized symmetric cumulants of charged pions $v_n$ $(n=2-4)$. 
We show that the normalized symmetric cumulants can differentiate the two EoSs while all other conditions remain the same. 
The conclusion does not change for various initial conditions and shear viscosity.
 %The MC Glauber modelis used to initialise the system with two different variants (a) wounded nucleons ($\varepsilon_{WN}$) and (b) binary collisions ($\varepsilon_{BC}$), for Au+Au  collisions at $\sqrt{s_{NN}}=200$ GeV and 62.4 GeV.  We found that the correlations $c(v_{2},v_{3})$ and $c(v_{3},v_{4})$ are always larger  for the first order phase transition than the cross over phase transition irrespective of the initial conditions. The same correlations were shown to be quite insensitive to the shear viscosity of the medium in a previous study by Niemi et al \cite{Niemi:2012aj}.
This indicates that these observables can be used to gain information about the 
QCD EoS from experimental data and can be used as an EoS meter. 
%Besides, we also found   $\sim 10\% -30\%$ difference in $c(\epsilon_{2}, v_{2})$ for the two different EoSs. These observations may be attributed to very different evolutionary dynamics of the system for the two different EoSs.%, as the speed  of sound becomes zero in first-order phase transition hence the  linear/non-linear coupling of $\epsilon_{n}$ and $v_{n}$.
\end{abstract}

\maketitle

%\tableofcontents

\section{\label{sec:Intro}Introduction}

It is well known that at low temperature and small baryon chemical potential, the degrees of freedom
of nuclear matter are color-neutral hadrons; whereas at high temperature or large baryon chemical potential,
the matter is in the form of quark-gluon plasma (QGP), in which the fundamental degrees of freedom are colored objects 
quarks and gluons.
The nuclear matter at high baryon density and finite temperature are believed to undergo first-order phase transitions, 
from the hadronic phase to the QGP phase, and the first-order phase transition line terminates at a critical point 
\cite{Datta:2012pj,Borsanyi:2010bp,Gavai:2004sd}. This is
because lattice QCD shows that the hadron to QGP transition is a crossover for vanishing baryon chemical 
potential at temperature $\sim$ 170 MeV \cite{Bernard:2004je,Bhattacharya:2014ara,Borsanyi:2010bp,Bazavov:2011nk}. 
For details on QCD phase diagram and critical points see review 
\cite{Fukushima:2010bq,Baym:2017whm}

Present theoretical models widely disagree with each other regarding the value of critical temperature and baryon chemical potential 
corresponding to the QCD critical point on the QCD phase diagram \cite{Schaefer:2007pw,Kovacs:2007sy}. 
Also, the existence of the QCD critical point is yet to be confirmed experimentally 
\cite{Aggarwal:2010wy,Adamczyk:2013dal,Adamczyk:2014fia,Adamczyk:2017wsl,Luo:2015ewa,Mohanty:2009vb}. 
It is crucial that phenomenologically 
motivated studies of the heavy-ion collision, such as relativistic hydrodynamics, be capable of accounting for the potential 
influence of such a critical point on experimental observables. Recently it was shown \cite{Pang:2016vdc}
that the goal mentioned above can be achieved by using relativistic hydrodynamic model, experimental data and 
a state-of-the-art deep-learning technique that uses a convolutional neural network to train the 
system.

\begin{figure*}
\begin{center}
%% Add the phase trajectory and momentum anisotropy plot for cross-over and phase transtn
 %\includegraphics[width =0.45\textwidth]{figures/EOS.pdf}
   \includegraphics[width =0.45\textwidth]{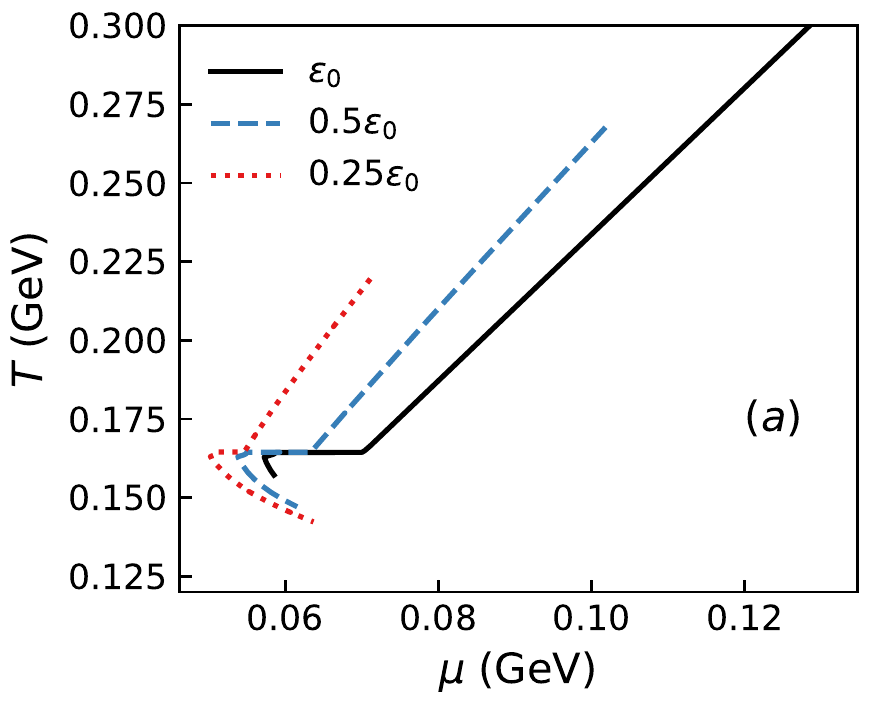}
    \includegraphics[width =0.45\textwidth]{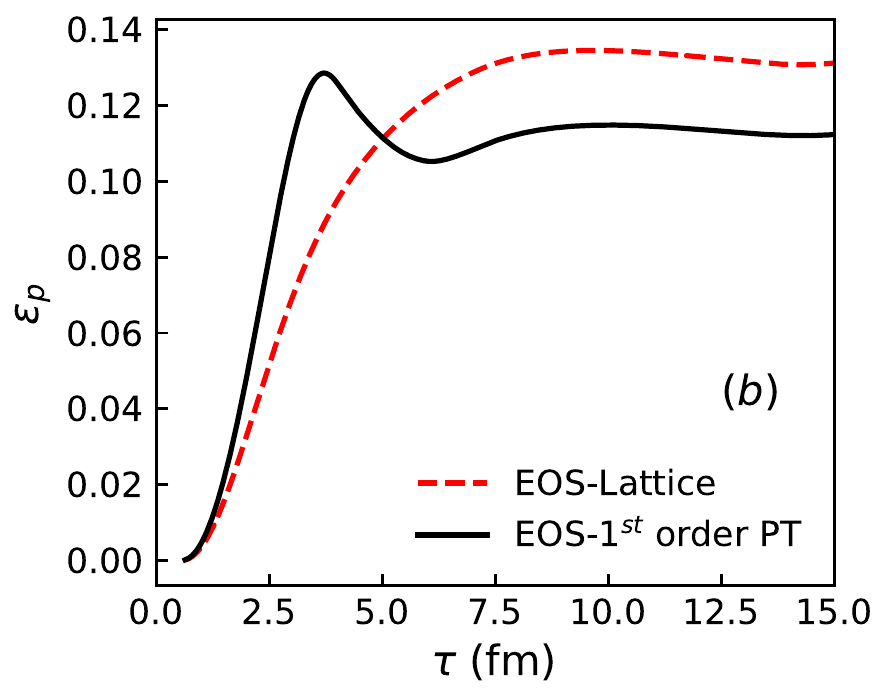}
  \end{center}
     \caption{(Color online) (a) Trajectories of different regions of the fireball during hydrodynamical evolution as a function temperature ($T$) and net baryon chemical potential ($\mu_B$) with 
  initial energy densities corresponding to 100$\%$, 50$\%$, 25$\%$ of the maximum energy density $\epsilon_0=16.2$~GeV/fm$^{3}$ till freeze-out 
  $\epsilon_F=0.3$~GeV/fm$^{3}$. The corresponding constant $s/n_B$ values are 156, 175 and 206 respectively. 
  (b) Time evolution of momentum anisotropy $\epsilon_p$ for cross-over (EoS Lattice) and
   first order phase transition ($1^{st}$ order PT) for Au-Au collisions at $\sqrt{s_{NN}}=62.4$ GeV respectively and impact parameter $b=8$~fm. 
  }
    \label{fig:epsTraj}
\end{figure*}

\begin{figure*}
    \centering
    \includegraphics[height =0.34\textwidth]{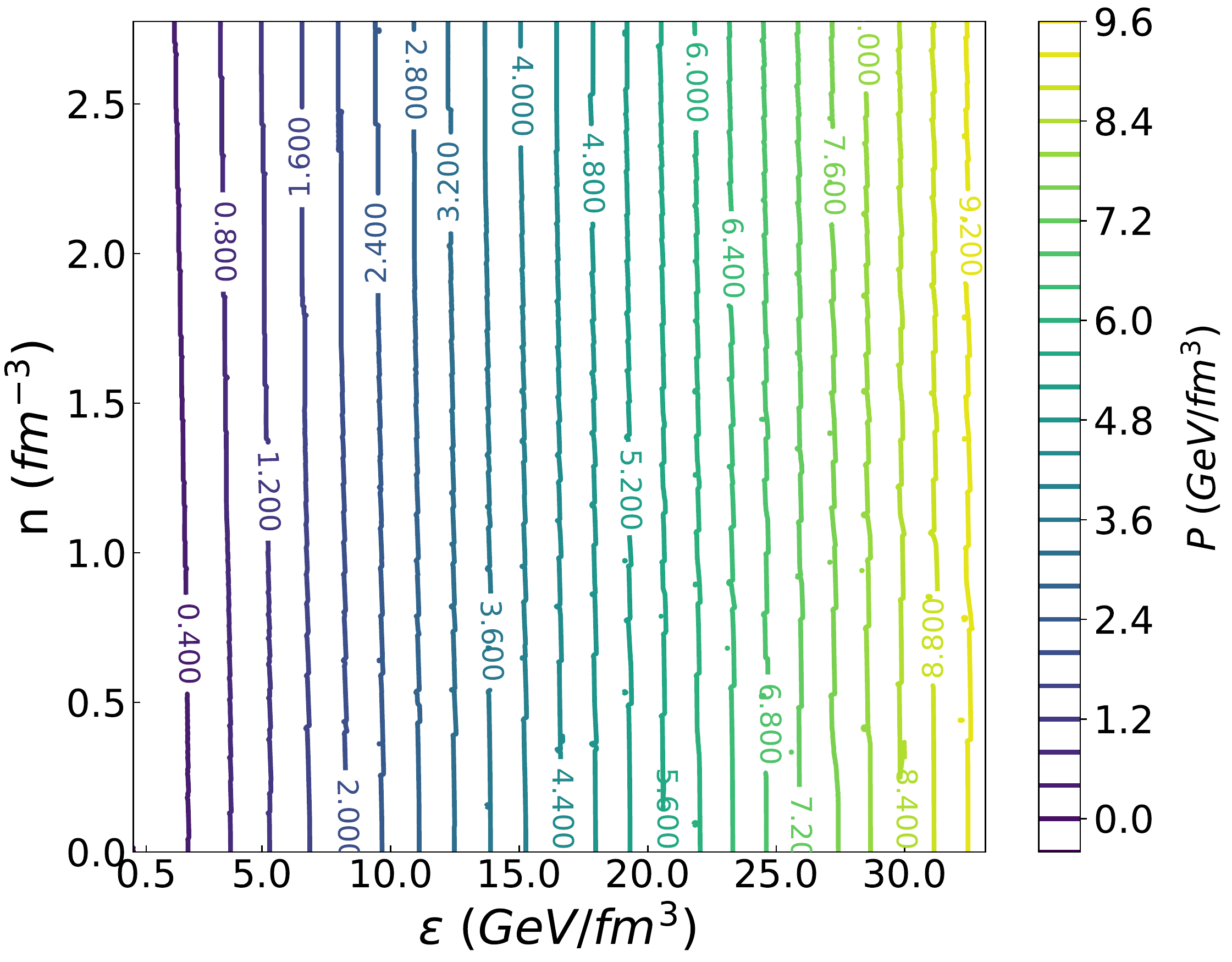}
        \includegraphics[height =0.34\textwidth]
        {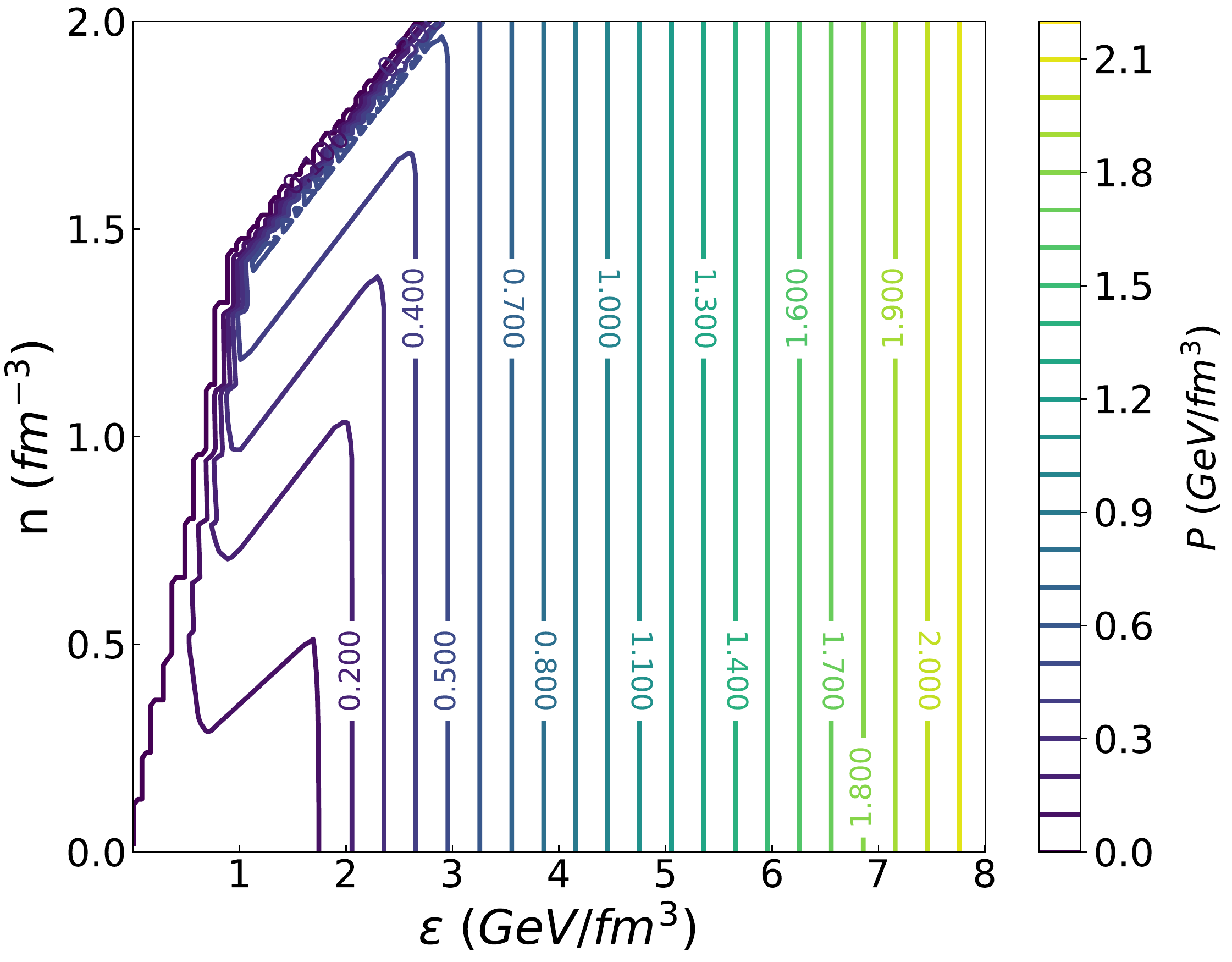}
     \caption{Equation of state $P(\varepsilon,n)$ shown as a contour plot in the $(\varepsilon,n)$ plane and the contours show lines of constant $P$. The left and right panel corresponds to crossover \cite{Noronha-Hostler:2019ayj} and first order  phase transition \cite{Baym:2017whm} respectively.}
    \label{fig:eos}
\end{figure*}
The present exploratory study aims to find a unique observable which connects QCD Equation of State (EoS) and the experimental data of heavy-ion collisions using one of the available dynamical models. We believe this effort will be complementary to the finding of \cite{Pang:2016vdc}.
It is well known that the initial energy/baryon densities follow 
very different space-time trajectories in the QCD phase diagram for crossover and first-order phase transitions \cite{Stephanov:1998dy}. 
This fact is shown in Fig.~\ref{fig:epsTraj}(a), the three different lines correspond to different entropy per baryon numbers (\(s/n_{B}\)), and
the trajectory for the crossover will follow \(\mu_{B}=0\) line. The first-order phase transition involves finite latent heat (kink in the trajectories), and the speed of sound is zero in the mixed-phase. We note that in the fluid dynamical picture, converting the initial spatial deformation to the final momentum anisotropy depends on the speed of sound (EOS) and other factors such as the viscosity of the medium. 
In the case of crossover, the speed of sound is never zero; the consequence of the two different EoS's can be seen in the temporal evolution of momentum space anisotropy \(\epsilon_{p}\) (defined later) shown in Fig.~\ref{fig:epsTraj}(b).
A different value of \(\epsilon_{p}\) corresponds to different elliptic flow. Although event averaged elliptic flow is sensitive
to the EoS used, it is also known that its value is suppressed in the presence of finite shear viscosity. Therefore the event-averaged elliptic flow is not a good indicator of the EoS.
Nevertheless, it is interesting to investigate what imprint of different
EoSs one can find in the correlation between the initial fluctuating geometry and the final flow coefficients in the event-by-event collisions \cite{Niemi:2012aj}.

We use the relativistic hydrodynamic model which has been very successful in simulating heavy-ion collisions and explaining experimental observables\cite{Kolb:2000sd,Nonaka:2000ek,Hirano:2001eu,Aguiar:2000hw,Chaudhuri:2006jd,Nonaka:2006yn,Song:2007ux,Baier:2006gy,Muronga:2006zw,Pratt:2008sz,Petersen:2008dd,Molnar:2009tx,Schenke:2010nt,Werner:2010aa,Holopainen:2010gz,Roy:2011pk,Bozek:2011ua,Pang:2012he,Akamatsu:2013wyk,Noronha-Hostler:2013gga,DelZanna:2013eua}. 
We find the linear/Pearson correlation (defined later) of initial geometric asymmetry to the corresponding flow coefficient 
(particularly the second-order flow coefficient \(v_{2}\)) is a unique observable which can differentiate 
between EoS with a first-order phase transition to that with a crossover transition irrespective of the 
initial condition used. It has been known that the event averaged \(v_{2},\) and the eccentricity
of the averaged initial state, \(\epsilon_{2}\) are approximately linearly correlated \cite{Niemi:2012aj,Plumari:2015cfa,Kolb:2003dz,Song:2008si},and the initial condition used \cite{Niemi:2012aj}.
Although the above finding sounds promising, we need to keep in mind that the initial eccentricities are not directly measurable in experiments; hence the correlation between the eccentricities and flow coefficients may not provide any practical information about the EoS's. 
But we can use the cumulants between flow coefficients such as $\mathrm{NSC}(v_{2},v_{3})$ and $\mathrm{NSC}(v_{3},v_{4})$ \cite{ALICE:2016kpq,Bilandzic:2013kga} which can always be measured experimentally without referring to any particular model. That is what motivates us to investigate these observables described above as an EoS meter. 

For the present study, we use a newly developed $2+1$-dimensional event-by-event viscous hydrodynamic code with 
an EoS \cite{Kolb:2000sd} with finite baryon chemical potential/density for low \(\sqrt{s_{NN}}\) collisions, and a lattice QCD+HRG
EoS \cite{Huovinen:2009yb} for the higher \(\sqrt{s_{NN}}\) collisions. There are ongoing efforts to 
construct EoS with a critical point \cite{Plumberg:2018fxo,Monnai:2019hkn}. Due to the present uncertainty in the location of 
QCD critical point, we refrain to use sophisticated EoSs.

The conservation equations are solved numerically by using the time-honored SHarp And Smooth Transport Algorithm
(SHASTA) \cite{Boris:1973tjt}. In the next section, we shall discuss the details of various tests performed to find the numerical accuracy of the code. As we will show, our code passes all the test cases satisfactorily.

The paper is organized as follows: in the next section, we discuss various aspects of the newly developed hydrodynamic code. In section \ref{sec:result} we show the comparison of experimental data of charged pion invariant yield and elliptic flow in Au+Au 200 GeV collisions to the simulation result for optical Glauber initial condition to test the code. In the same section, we also discuss the correlation between different observable and the effect of EoS on them. 
Finally in section \ref{sec:Conclusion} we conclude and discuss some future possibilities.
Throughout this article we adopt the units \(\hbar=c=k_{\mathrm{B}}=1 .\) The signature of the
metric tensor is always taken to be \(g^{\mu \nu}=\operatorname{diag}(+1,-1,-1,-1) \).
Upper greek indices correspond to contravariant and lower greek indices covariant. The three vectors are denoted with Latin indices.

%%%%%%%%%%%%%%%%%%%%%%%%%%%%%%%%%%%%%%%%%%%%
%%%%        SECTION- -------> FORMALISM , V Roy 31st July, 2019
%%%%%%%%%%%%%%%%%%%%%%%%%%%%%%%%%%%%%%%%%%%%

\subsection{Conservation equations}
Relativistic hydrodynamics model of high-energy heavy-ion collisions assumes that matter
created in collisions  reaches a state of local thermal equilibrium at time $\tau_0$. 
The evolution of the thermalised nuclear matter is governed by the conservation equation of energy-momentum tensor and net baryon current,
\begin{eqnarray}\label{Eq:EMDen}
 \partial_{\mu}T^{\mu\nu}&=&0,\\\label{Eq:NumDen}
 \partial_{\mu}J^{\mu}&=&0,
\end{eqnarray}
where the the energy-momentum tensor and the net baryon current can be expressed as
\begin{eqnarray}
 T^{\mu\nu}&=&(\epsilon+P)u^{\mu}u^{\nu}-Pg^{\mu\nu}+\pi^{\mu\nu},\\
 J^{\mu}&=&nu^{\mu}.
\end{eqnarray}
Here $\epsilon$ is the local energy density, $P$ is pressure, $g^{\mu\nu}$ is the metric tensor, $n$ is the net baryon density, $u^{\mu}$ is the time-like 4-velocity with $u^{\mu}u_{\mu}=1$ and $\pi^{\mu\nu}$ is the shear-stress tensor. The evolution equation for the shear-stress tensor is given as \cite{Denicol:2012cn,ISRAEL1976213}
\begin{equation}
\begin{aligned}
\Delta_{\alpha \beta}^{\mu v} \tau_{\pi} D \pi^{\alpha \beta}+\pi^{\mu v}=& 2 \eta \sigma^{\mu v}-\frac{4}{3} \tau_{\pi} \pi^{\mu v} \theta-\frac{10}{7} \tau_{\pi} \Delta_{\alpha \beta}^{\mu \nu} \sigma_{\lambda}^{\alpha} \pi^{\beta \lambda} \label{Eq:Shear}
\end{aligned}
\end{equation}
where $\eta$ is the shear viscosity coefficient, $D=u^{\mu} \partial_{\mu}$ is the comoving time derivative, $\sigma^{\mu v}=\Delta_{\alpha \beta}^{\mu v} \partial^{\alpha} u^{\beta}$ is the shear tensor, $\theta=\partial_{\mu} u^{\mu}$ is the expansion rate, and $\Delta_{\alpha \beta}^{\mu v}=\left(\Delta_{\alpha}^{\mu} \Delta_{\beta}^{v}+\right.$
$\left.\Delta_{\alpha}^{v} \Delta_{\beta}^{\mu}-2 / 3 \Delta^{\mu v} \Delta_{\alpha \beta}\right) / 2$, with $\Delta^{\mu v}=g^{\mu v}-u^{\mu} u^{v}$. The transport coefficients in the above equation were obtsined in the massless limit with the relaxation time $\tau_\pi$ being $\tau_\pi=5\eta/(\epsilon+P)$.

In the present work, we will be using two kinds of  EoS's:\\
%% Give Ref. to Heinz anslo some desrpn about finite muB
(i) A parameterized EoS \cite{Noronha-Hostler:2019ayj} shown in Fig.~\ref{fig:eos} (left panel) which has a cross-over transition between high temperature QGP phase obtained from lattice QCD and a hadron resonance gas below the crossover temperature.  In order to have  a smooth pressure profile as a function of $\varepsilon$ and $n$, the pressure is first expanded as a as a Taylor series in powers of $\mu_B/T$. The associated Taylor coefficients which are known form LQCD \cite{Borsanyi:2013bia,Bazavov:2014pvz} are parameterized using polynomial of ninth order.
(ii) A  parameterized EoS \cite{Kolb:2000sd} with a first order phase transition at finite net baryon number density shown in Fig.~\ref{fig:eos} (right panel). The EoS connects a non-interacting massless QGP gas at high temperature to a hadron resonance gas, of masses up to 2 GeV, at low temperatures through a first order phase transition.  The bag constant $B$ is a parameter adjusted to $B^{1/4}=230~$MeV, to yield a critical temperature of $T_c=164~$MeV. We note that the choice of bag parameter used here is not unique, 
it may vary between  $B^{1/4}=150-300 $ MeV \cite{Baym:2017whm}. Both of these EoS will be used in the  hydrodynamic simulation  corresponding to \(\sqrt{s_{\mathrm{NN}}}=\) 62.4 GeV collisions at a finite net baryon density.\par

Instead of solving the full $3+1-$ dimensional hydrodynamics equations, we consider a simplified evolution by assuming boost invariant expansion in the $z$ direction. This can be done easily by working in the Milne coordinates with metric tensor given by $g^{\mu\nu}=(1,-1,-1,-1/\tau^2)$. The velocity 4-vector in this coordinate (physical quantities in this coordinate is written with a tilde sign), $\tilde U^{\mu}=\tilde\gamma(1,{\tilde v_{x}},{\tilde v_{y}},0)$, where $\tilde v_{i}=v_i\cosh\eta,\hspace{0.1cm}(i=x,y)$ and 
$\tilde\gamma=1/\sqrt{1-\tilde v_{\bot}^2}$ with $\tilde v_{\bot}^2={\tilde v_{x}}^2+{\tilde v_{y}}^2$ . \par

Conservative equations of the form Eqs.~(\ref{Eq:EMDen},\ref{Eq:NumDen},\ref{Eq:Shear}) can be solved accurately using flux-corrected transport (FCT) algorithms Refs.~\cite{Boris1993LCPFCTAFT,POWELL1999284,Marti:1999wi}
without violating the positivity of mass and energy, particularly near shocks and other discontinuities. Here, we use the SHASTA ("SHarp and Smooth Transport Algorithm"),  which is a designed to solve partial differential equations of the form 
\begin{equation}
 \partial_t(A)+\partial_x(v_x A)+\partial_y(v_y A)=B(t,x,y)
\end{equation}
where $A=A(t,x,y)$ is for example $T^{00},T^{0i},..,,v_i$ is the $i^{\mathrm{th}}$ component of three-velocity, and $B(t,x,y)$ is a source term. The local rest frame charge and energy densities 
are given as
\begin{eqnarray}
 n&=&j^0\sqrt{1-\tilde v_{\bot}^2},\\
 \epsilon&=&T^{00}-(\tilde v_xT^{0x}+\tilde v_yT^{0y}),
\end{eqnarray}
while the velocity components are calculated using an one-dimensional root search algorithm, by
iterating the following transcendental equation
\begin{equation}
 \tilde v_{\bot}=\frac{{T^{\tau\bot}}}{T^{\tau\tau}+P(\epsilon)},
 \end{equation}
 such that the components of velocity are given by
 \begin{equation}
 \tilde v_i=\tilde v_{\bot} \frac{T^{\tau i}}{T^{\tau\bot}}
\end{equation}
where $T^{\tau\bot}=\sqrt{{(T^{\tau x})}^2+{(T^{\tau y})}^2}$. In this velocity finding algorithm 
we keep the accuracy $\sim 10^{-22}$.

We have tested our numerical simulation against $(1+1)-$ dimensional Riemann problem which has an analytical solution in the perfect-fluid limit. The numerical solution reproduces the analytic solution with a very high precision with sufficiently high numerical resolution. We have similarly checked and verified the applicability of our numerical simulation to $2+1-$ dimension by having the initial discontinuity in the $1+1-$ dimensional Riemann problem in the plane $y=-x$. For small viscosities $\eta/s=10^{-2}$, we have verified that the viscous version of our simulation reproduces the result of our simulation with $\eta/s=0$ but with anti-diffusion coefficient $A_\mathrm{ad}=0.9$, which is the numerical analog of physical viscosity. Finally, we have checked that our numerical simulation reproduces Gubser flow with initial condition fixed such that, the system expands and cools down much faster than typical heavy-ion collisions.
In the numerical code we use the CORNELIUS  subroutine reported in \cite{Huovinen:2012is}. 
The subroutine uses an improved version of the original Marching Cube algorithm and extends the different
distinct possible topological configuration from $15$ to $33$ and thus creates a consistent surface with lesser holes
 and no double counting which is required for event-by-event hydrodynamics study.

\section{\label{sec:result}Result}
\subsection{Testing for smooth Glauber}
Before we show the result of event-by-event hydrodynamics simulation we show here the results from smooth optical Glauber 
model for $\sqrt{s_{NN}}=200$, $62.4$~GeV collision energies respectively. For all the calculations, the spatial extension of 
the numerical grid is set to $36\times 36$ fm$^2$. The spatial grid spacing is set to $\Delta x=\Delta y=0.09$ fm and the temporal 
spacing is set to $\Delta t=0.04$ fm. The initial time for all collisional erergies is fixed to $\tau_0=0.6$~fm.

\begin{figure}
\label {fig:smthSpctra}
    \begin{center}
    %\centering
    \includegraphics[width =0.35\textwidth]{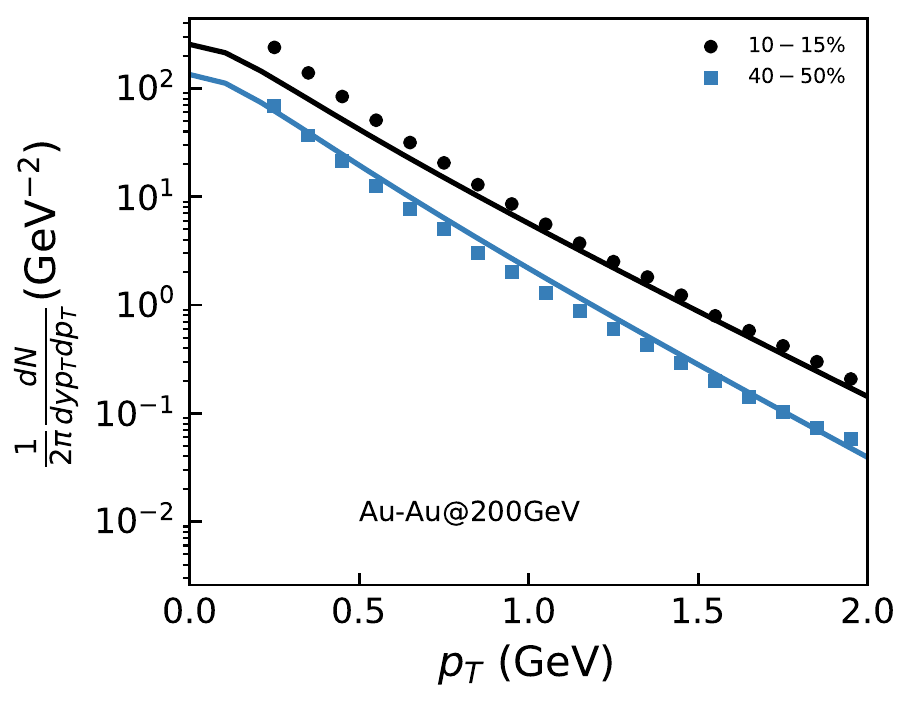}
    \includegraphics[width =0.35\textwidth]{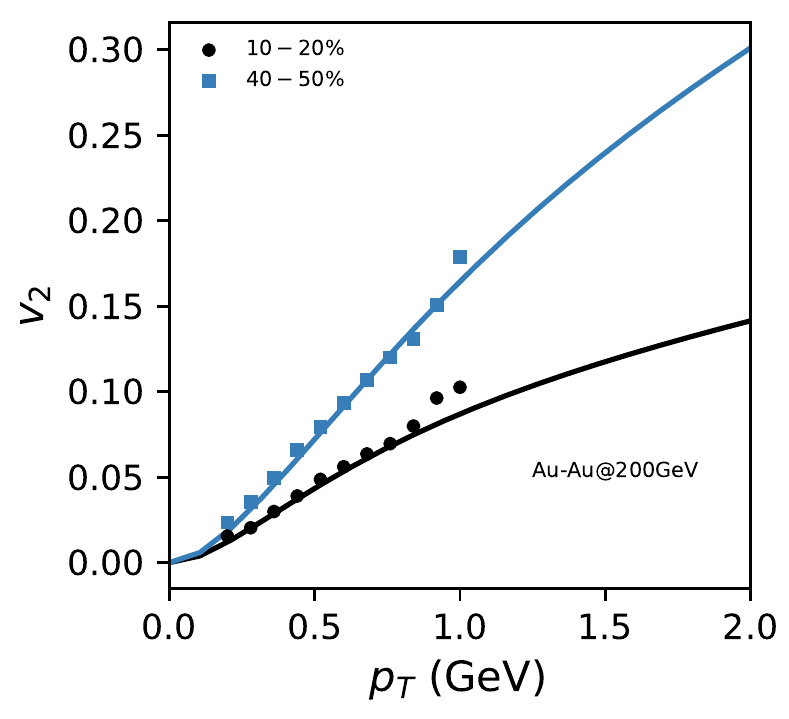}
    \end{center}
    \caption{(Color online) (a) Top panel : Comparison of experimentally measured invariant yield of $\pi^{-}$  to 
    the result obtained from ideal hydrodynamic simulation for Au+Au collisions at $\sqrt{s_{NN}}=200$ GeV collisions. 
    Experimental and simulation result for $10\%-15\%$ collision centrality is shown by black circles and black line
    respectively.  The same for   $40\%-50\%$ collision centrality is shown by blue square and blue line respectively.  
    (b) Bottom panel: same as top panel but for elliptic flow $v_{2}$ of $\pi^{-}$.} 
    \label{fig:spectrav2}
\end{figure}

For Au-Au collisions at $\sqrt{s_{NN}}=200$~GeV we use EoS-Lattice shown in Fig.~\ref{fig:eos}(a) with $\mu_B=0$ MeV and the 
freeze-out energy density is taken to be $\varepsilon_F=0.28$~GeV/fm$^{3}$ which corresponds to a temperature $T_f=137$~MeV. 
The central energy density is fixed to 
$\varepsilon_0=45$~GeV/fm$^{3}$ for $10\%-15\%$ collision centrality which explains the corresponding experimental data. 
The top panel of Fig.~\ref{fig:spectrav2} shows the comparison of invariant yield  of $\pi^{-}$ for $10\%-15\%$ and $40\%-50\%$ centrality collisions. 
Experimental data measured by PHENIX collaboration \cite{Adler:2003cb} are shown by different symbols and the corresponding simulation results are shown 
by lines. We found reasonable agreement with the experimental data except at low momentum $p_{T} < 0.5 $ GeV, this is because we neglected the 
contribution of $\pi^{-}$ yield coming from various heavier resonance decay. We plan to incorporate the resonance 
decay in a later version of the code. The comparison of simulated elliptic flow with the experimental data of STAR collaboration \cite{Adams:2004bi} are shown in the bottom panel 
of Fig.~\ref{fig:spectrav2}. We can see that the simulated result nicely explain the observed asymmetry in the final momentum spectra of $\pi^{-}$. 

For Au-Au collisions at $\sqrt{s_{NN}}=62.4$~GeV we use EoS-$1^{st}$ order PT  shown in Fig.~\ref{fig:eos}(b) and the 
freeze-out energy density is fixed to $\varepsilon_F=0.3$~GeV/fm$^{3}$. The initial central energy density is set to 
$\varepsilon_0=16$~GeV/fm$^{3}$ for $0\%-15\%$ collision centrality. Similarly, the initial central net baryon density is fixed to $n_0=0.4$~fm$^{-3}$.
We have checked that the above parameters explains the invariant yield of $\pi^{-}$\cite{Back:2006tt} across
various centralities. In Fig.~\ref{fig:epsTraj}(a) we have plotted the trajectories of different regions of the fireball during hydrodynamical evolution as a 
function temperature ($T$) and net baryon chemical potential ($\mu_B$) with initial energy densities corresponding to 100$\%$, 50$\%$, 25$\%$ of the maximum energy
density $\epsilon_0=16.2$~Gev/fm$^{3}$ till freeze-out $\epsilon_F$. The corresponding constant $s/n_B$ for the above lines are 156, 175 and 206 respectively. 
Momentum space anisotropy $\epsilon_p$ defined as
\begin{equation}
 \epsilon_p=\frac{\int d^2x \left(T^{xx}-T^{yy}\right)}{\int d^2x \left(T^{xx}+T^{yy}\right)}
\end{equation}
where $T^{ii}$ are components of energy-momentum tensor $T^{\mu\nu}$ are plotted in Fig.~\ref{fig:epsTraj}(b) both for EoS-Lattice and first order phase transition at $b=8$ fm.
This completes the testing of our code. In the next section, we shall study the effect of phase transition on the various flow correlation
which is the main goal of this paper.

%%%%%     Figure -1 , Spectra All centrality %%%%%%%%%%%%%%%%%%

%%%%%%%%%%%%%%%%%%%%%%%%%%%%%%%%%%%%%%%%%%%%%%%%%
%%%%%%%%%.    MC GLAUBER and FLow correlation %%%%%%%%%%%%%%%%%%%%

\subsection{Normalised symmetric cumulant and event-by-event flow correlation}
Flow observables obtained from multiparticle correlations, symmetric cumulants (SC), are introduced, which are defined as \cite{ALICE:2016kpq,Bilandzic:2013kga}
\begin{widetext}
\begin{equation}\label{Eq:SC}
\begin{aligned}
\mathrm{SC}(m, n) \equiv\left\langle\left\langle\cos \left(m \varphi_{1}+n \varphi_{2}-m \varphi_{3}-n \varphi_{4}\right)\right\rangle\right\rangle_{c}=&\left\langle\left\langle\cos \left(m \varphi_{1}+n \varphi_{2}-m \varphi_{3}-n \varphi_{4}\right)\right\rangle\right\rangle \\
&-\left\langle\left\langle\cos \left[m\left(\varphi_{1}-\varphi_{2}\right)\right]\right\rangle\right\rangle\left\langle\left\langle\cos \left[n\left(\varphi_{1}-\varphi_{2}\right)\right]\right\rangle\right\rangle \\
=&\left\langle v_{m}^{2} v_{n}^{2}\right\rangle-\left\langle v_{m}^{2}\right\rangle\left\langle v_{n}^{2}\right\rangle
\end{aligned}
\end{equation}
\end{widetext}
with the condition $m \neq n$ for two positive integers $m$ and $n$. The double angular brackets indicate that the averaging procedure has been performed in two steps- first over all distinct particle quadruplets in an event, and then in the second step the single-event averages were weighted with number of combinations. The four-particle cumulant in Eq.~\eqref{Eq:SC} is less sensitive to non-flow correlations than any 2- or 4-particle correlator on the right-hand side taken individually \cite{Borghini:2000sa,Borghini:2001vi}. The last equality is true only in the absence of non-flow effects. The observable in Eq.~\eqref{Eq:SC} is zero in the absence of flow fluctuations, or if the magnitudes of harmonics $v_m$ and $v_n$ are uncorrelated. Experimentally it is more reliable to measure the higher order moments of flow harmonics $v^k_n$ $(k\geq2)$ with 2- and multiparticle correlation techniques, than to measure the first moments $v_n$ with the event plane method, due to systematic uncertainties involved in the event-by-event estimation of symmetry planes.

One can also define normalised symmetric cumulants (NSC) as,
\begin{equation}
\operatorname{NSC}(m, n) \equiv \frac{\operatorname{SC}(m, n)}{\left\langle v_{m}^{2}\right\rangle\left\langle v_{n}^{2}\right\rangle}.
\end{equation}
Normalized symmetric cumulants reflect the strength of the correlation between $v_{m}$ and $v_{n}$, while $\mathrm{SC}(m, n)$ has contributions from both the correlations between the two different flow harmonics and the individual harmonics. In the present work, we calculate the symmetric cumulants using the last equality in Eq.~\eqref{Eq:SC}, due to the absence of non-flow effects in our hydrodynamic formulation.   

The hydrodynamics equation of motion has to be initialized at an initial time $\tau=\tau_0$, by specifying the initial energy densities $\varepsilon(\tau_0,x,y)$ or the entropy densities $s(\tau_0,x,y)$ and the flow velocities $u^{\mu}$ of a relativistic fluid.  Two of such models that we are going to use are the Glauber model \cite{Miller:2007ri} and the TRENTo model \cite{Moreland:2014oya}.

Given a pair of projectiles labeled $A$ and $B$ collide along the beam axis $z$ and let $\rho_{A,B}(x,y,z)$ be the density of the nuclear matter that participate in inelastic collisions. $\rho(x,y,z)$ is usually given by the Woods-Saxon profile, given as
\begin{equation}
\rho(x,y,z)=\frac{\rho_{0}}{1+\exp [(r-R) / a]},
\end{equation}
where $\rho_0, R$ and $a$ are the normalization, size of the nucleus, and the stiffness of the edge of nucleon distribution profile respectively while $x,y,z$ are the spatial coordinates. Each  projectile may then be represented by its participant thickness, given by
\begin{equation}
 T_{A,B}=\int dz \rho_{A,B}(x,y,z).
\end{equation}
We shall assume  that their exists a function $f(T_A,T_B)$ which converts the projectile thickness to entropy production, namely $f\propto ds/dy|_{\tau=\tau_0}$. In a two-component Glauber model the function is the sum of $f\sim T_A +T_B$, which is proportional to the number of \textit{wounded} nucleons $N_{WN}$ and a quadratic term $f\sim T_AT_B$, which is proportional to the number of \textit{binary} collisions $N_{BC}$. The complete function $f$ is given as
\begin{equation}
 f\sim (T_A +T_B) + \alpha T_A T_B.
\end{equation}
The proportionality constants and the other relevant details can be found in \cite{Miller:2007ri}. However, in general the energy density profile in the reaction zone of nucleus-nucleus collision fluctuates from event to event due to the quantum fluctuations of the nuclear wave function. This fluctuations are attributed to the fluctuations in the positions of participating nucleons. In such a case, the previous decribed smooth Glauber distribution will then be an ensemble average of a large number of fluctuating initial distribution. This can be achieved by extending the smooth Glauber model to their Monte Carlo (MC) versions. These fluctuating initial distributions breaks the rotational and reflection symmetry of the smooth distributions.

The TRENTo model  \cite{Moreland:2014oya} for the initial condition assumes the a scale invariant form of the function $f$, i.e.,
\begin{equation}
 f(cT_A,cT_B)=cf(T_A,T_B),
\end{equation}
for any nonzero constant $c$. One can clearly see that the above is broken by the binary collision term $\alpha T_A T_B$ in the Glauber model. TRENTo assumes a reduced thickness function given by
\begin{equation}
f=T_{R}\left(p ; T_{A}, T_{B}\right) \equiv\left(\frac{T_{A}^{p}+T_{B}^{p}}{2}\right)^{1 / p}.
\end{equation}
Various limiting cases of the function $f$, for different choices of parameter $p$ can be found in \cite{Moreland:2014oya}. In the present study we assume $p=0$, in which case $f=\sqrt{T_AT_B}$, which is the geometric mean of the $T_A$ and $T_B$. We will be using the above two models in order to see the sensitivity of various correlations to the initial conditions used for the simulations.

\begin{figure*}
\begin{center}
 \includegraphics[width =0.40\textwidth]{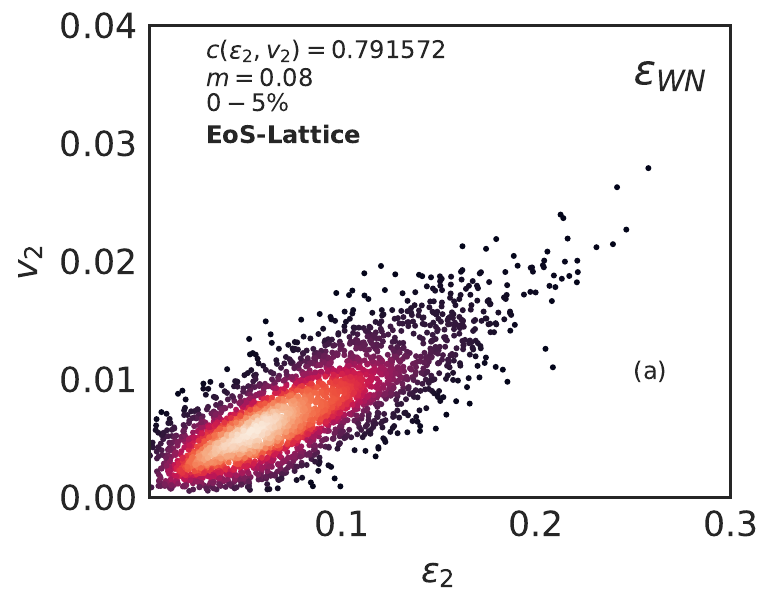}
\includegraphics[width =0.40\textwidth]{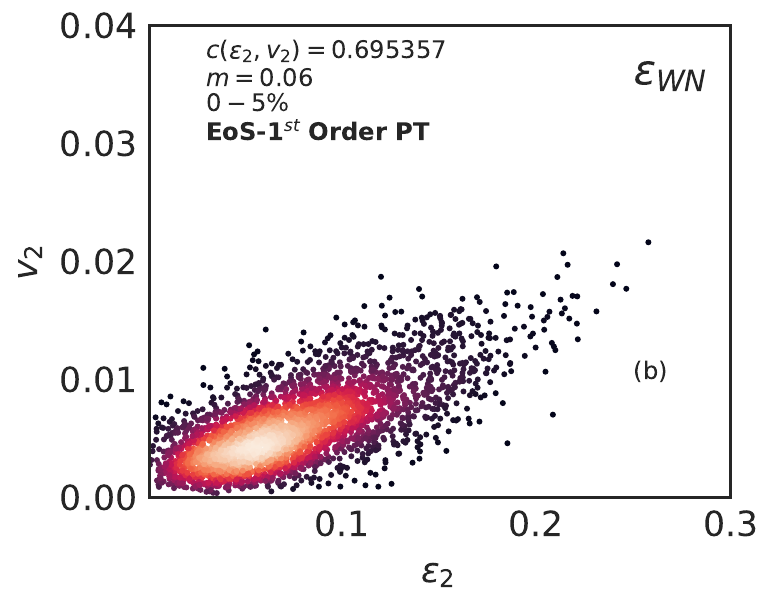}
\includegraphics[width =0.40\textwidth]{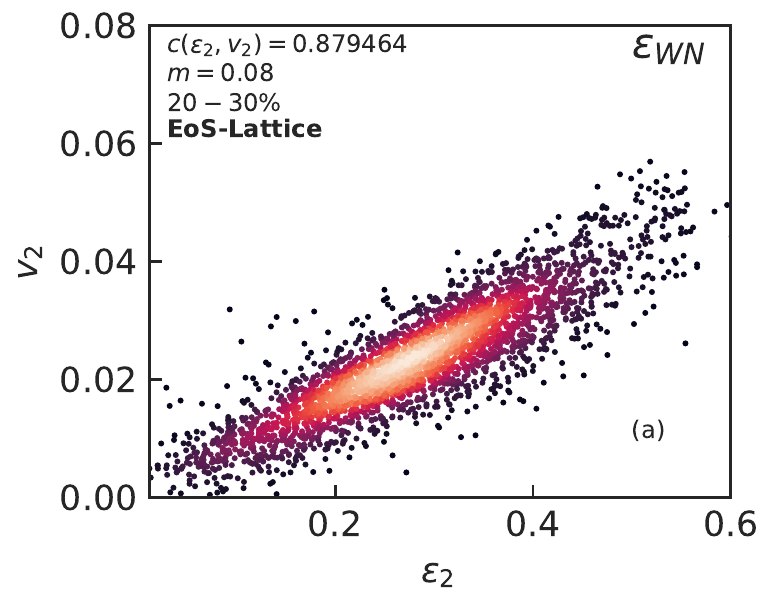}
\includegraphics[width =0.40\textwidth]{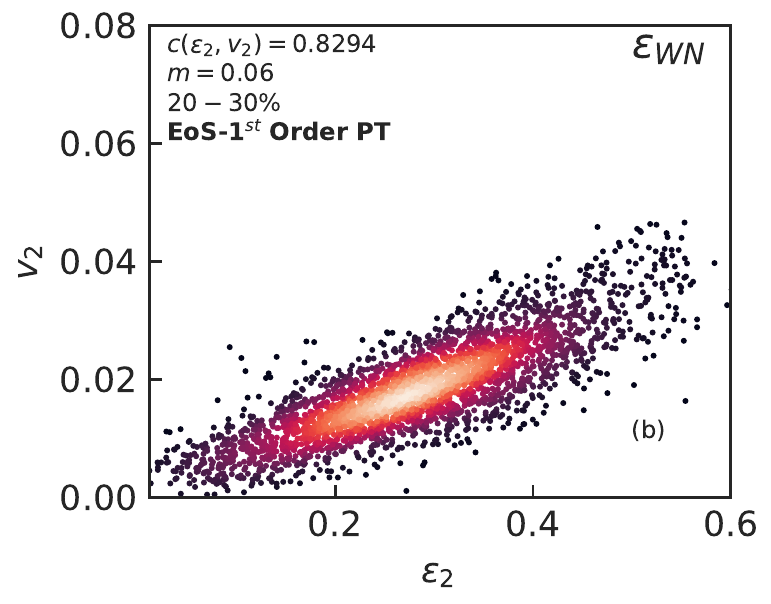}
\includegraphics[width =0.40\textwidth]{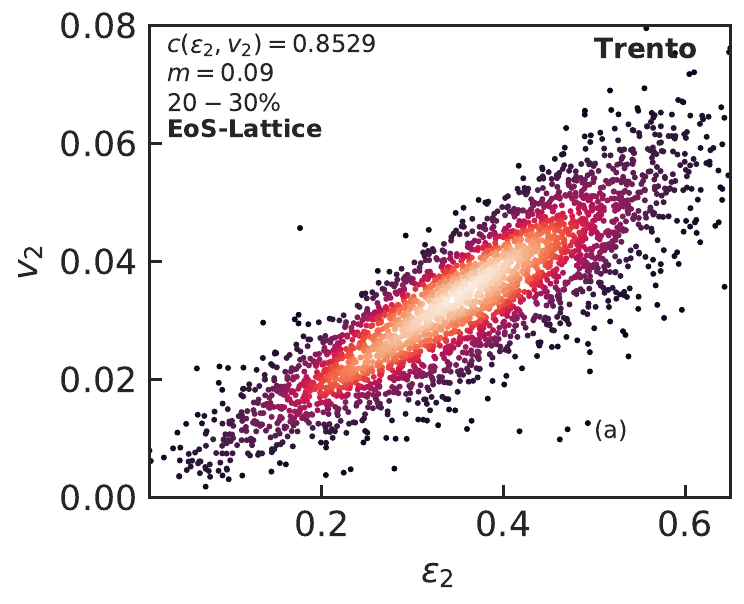}
\includegraphics[width =0.40\textwidth]{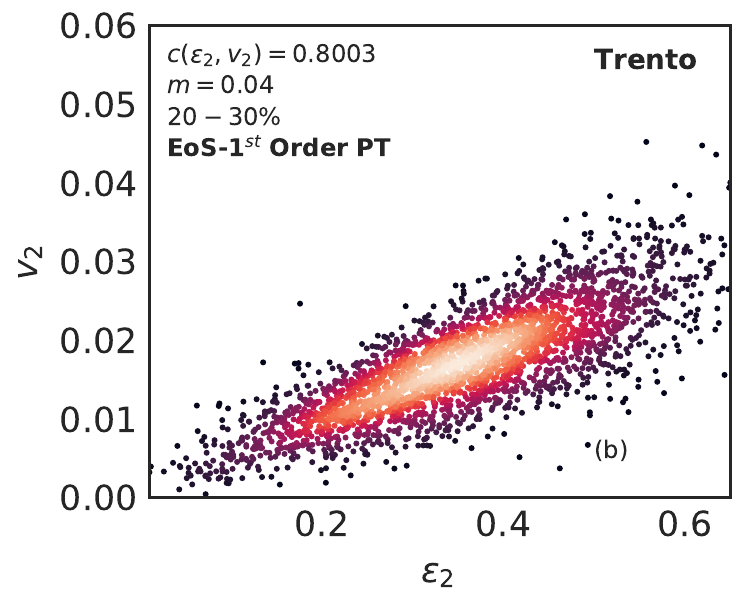}
\end{center}
\caption{ (Top row) Event-by-event distribution of $v_{2}$ vs $\epsilon_{2}$  for $0\%-5\%$ Au+Au collisions at $\sqrt{s_{NN}}=62.4$ GeV. The initial energy density and $\epsilon_{2}$ is obtained from MC-Glauber model. (Middle row) Same as top row but for $20\%-30\%$ centrality. (Bottom row) Initial conditions from TRENTo model at $20\%-30\%$ centrality. The left column (a) is for crossover transition while the right column is for first order phase transition.}
\label{fig:v2e2}
\end{figure*}

 \begin{figure*}
\centering
\includegraphics[width =0.48\textwidth]{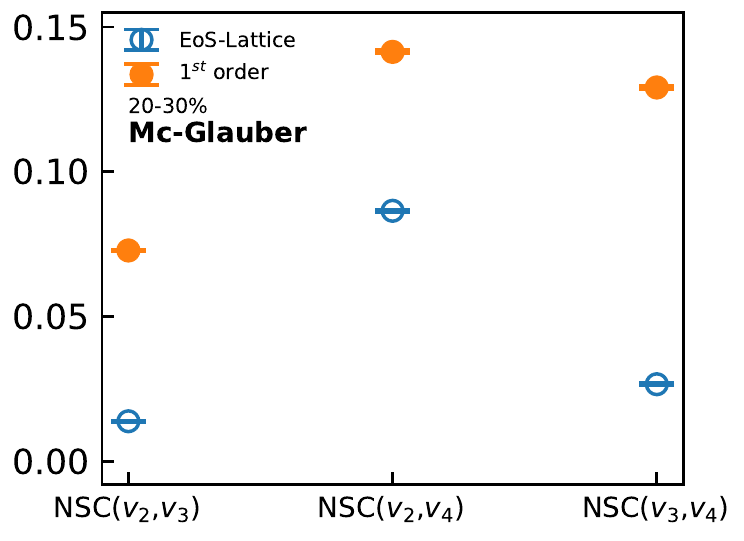}
\includegraphics[width =0.48\textwidth]{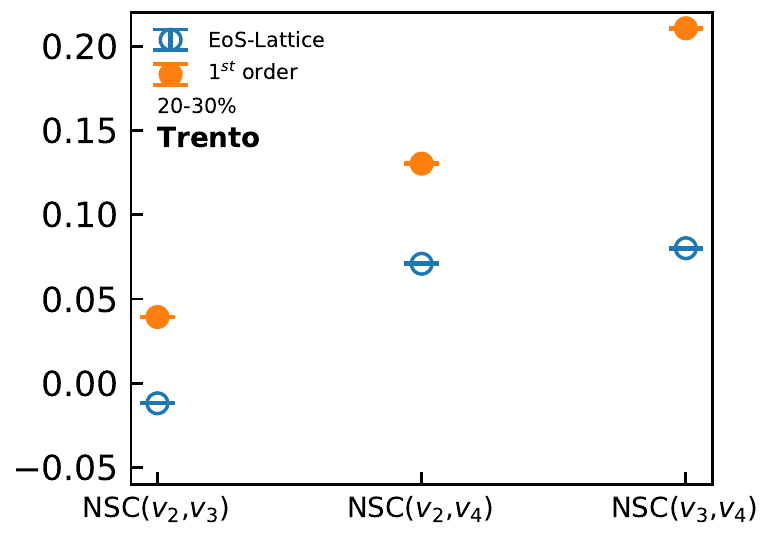}
\caption{ (Left panel) Normalized symmetric cummulants $\mathrm{NSC}(m,n)$ for EoS-Lattice (solid orange circles), and first order phase transition (open blue circle) for $20\%-30\%$ collision centrality. The initial energy density is obtained from wounded nucleons $(\varepsilon_{WN})$ in MC-Glauber model. (Right panel) Same as left panel but for TRENTo initial conditions.}
\label{fig:diffvnvm2030}
\end{figure*}
 \begin{figure*}
\centering
\includegraphics[width =0.48\textwidth]{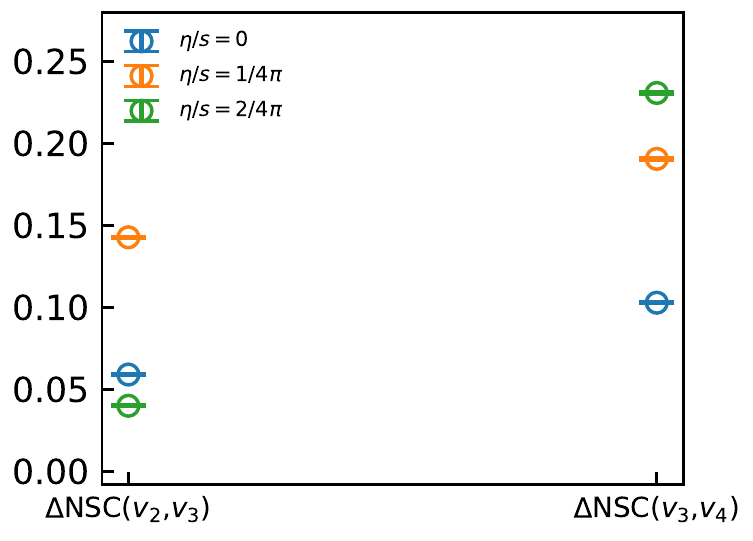}
\caption{ The difference between the $\mathrm{NSC}(m,n)$ calculated for EoS- with first order phase transition and that with crossover for different values of $\eta/s$. The initial energy density is obtained from wounded nucleons $(\varepsilon_{WN})$ in MC-Glauber model for $20\%-30\%$ collision centrality.}
\label{fig:diffvnvm2030visc}
\end{figure*}
We are interested in the study of the difference in fluid dynamical response of the system to the initial geometry (anisotropy) for two different EoS. In the literature this initial geometry/anisotropy of the overlap zone of two colliding nucleus is quantified in terms of eccentricities \(\epsilon_{n} \) \cite{Alver:2010gr,Alver:2010dn}:
\begin{equation}\label{Eq:eccen}
\epsilon_{n} e^{i n \Phi_{n}}=-\frac{\int d x d y r^{n} e^{i n\phi} \varepsilon\left(x, y, \tau_{0}\right)}{\int d x d y r^{n} \varepsilon\left(x, y, \tau_{0}\right)}.
\end{equation}
where $r^{2}=x^{2}+y^{2}, \phi$ is the spatial azimuthal angle, and $\Phi_{n}$ is the participant angle given by,
\begin{equation}
\Phi_{n}=\frac{1}{n} \arctan \frac{\int d x d y r^{n} \sin (n \phi) \varepsilon\left(x, y, \tau_{0}\right)}{\int d x d y r^{n} \cos (n \phi) \varepsilon\left(x, y, \tau_{0}\right)}+\pi / n.
\end{equation}
$\epsilon_n$ is basically the eccentricity of a polygon of $n^{\mathrm{th}}$ order, which can be reconstructed from a initial distribution generated from a MC Glauber or TRENTo model in a given event. Such a $n^{\mathrm{th}}$ order initial distribution generates a $n^{\mathrm{th}}$ order harmonic flow $v_n$ analogous to $v_2$.

As described before, the azimuthal momentum anisotropy is characterized in
terms of the coefficients $v_{n}$ of the Fourier expansion of the
single particle azimuthal distribution. We use the following definition to calculate $v_{n}$ in each event:
\begin{equation}\label{Eq:vndef}
 \frac{d^3N}{p_Tdp_Td\phi dy}=\frac{d^2N}{2\pi p_Tdp_Tdy}\sum_{n=-\infty}^{n=\infty}v_n(y,p_T)e^{i n\left(\phi-\Psi_n(y,p_T)\right)}.
\end{equation}

In Eq.~\eqref{Eq:vndef} $v_n$ can be calculated as the expectation $v_n = \langle \cos(n\phi - n\Psi_n)\rangle$, with respect to the associated event plane angle $\Psi_n$ of $n$-th order harmonics and is defined as
\begin{equation}
 \Psi_n=\frac{1}{n}\arctan\left(\frac{\langle p_T \sin n\phi\rangle}{\langle p_T \cos n\phi\rangle}\right).
\end{equation}
The angle brackets indicates the averaged value with respect to the particle spectrum.

Event-by-event hydrodynamics is one of the most natural way to model azimuthal momentum anisotropies $v_n$ Eq.~\eqref{Eq:vndef} generated by fluctuating initial-state anisotropies $\epsilon_n$ Eq.~\eqref{Eq:eccen}, which are generated by the highly fluctuating initial conditions in experiments. The largest source of uncertainty in these hydrodynamic models are the initial conditions. Since, direct measurement of bulk properties of matter like EoS in experiments is not possible, we try to identify hydrodynamic responses which in one hand can be calculated in experiments and in other hand are also tolearant to uncertainities in the model parameters viz., the initial conditions, shear viscosity etc. 

It has been known that the event averaged $v_{n}$, and the eccentricity of the averaged initial state, $\epsilon_{n}$ are approximately linearly related \cite{Kolb:2003dz,Song:2008si}  for $n<4$ but the same may not be true for higher-order flow coefficients. It is also known that the same linear relationship holds approximately between 
$\epsilon_{n}$ and $v_{n}$  even for event-by-event fluctuating conditions \cite{Niemi:2012aj}. Here we first study how the event-by-event correlation between  $\epsilon_{n}$ and $v_{n}$  is changed when the system 
undergoes either a first phase transition or a cross over. In order to quantify the  linear correlation we use Pearson's correlation coefficient which is defined as
\begin{equation}
\label{eq:correlation}
c(x, y)=\left\langle\frac{\left(x-\langle x\rangle_{\mathrm{ev}}\right)\left(y-\langle y\rangle_{\mathrm{ev}}\right)}{\sigma_{x} \sigma_{y}}\right\rangle_{\mathrm{ev}},
\end{equation}
where $\sigma_x$ and $\sigma_y$ are the standard deviations of the quantities $x$ and $y$. 
The correlation coefficient ranges from $-1$ to $1$. A value of $1(-1)$ implies that a linear (anti-linear) 
correlation between $x$ and $y$.  A value of $0$ implies that there is no linear correlation between the variables.
In a previous study, it was shown that the Pearson correlator $c(\epsilon_{2},v_{2})$ is  almost insensitive to 
the different initial condition and  the value of shear viscosity over entropy density of the fluid \cite{Niemi:2012aj}.
Further, assuming an approximate linear relationship  between the $\epsilon_{2}$ and $v_{2}$, we can write for event-by-event case
 \begin{equation}
 \label{eq:linv2e2}
 v_{2}= m \epsilon_{2}+\delta,
 \end{equation}
where \(m=\left\langle v_{2}\right\rangle_{\mathrm{ev}} /\left\langle\epsilon_{2}\right\rangle_{\mathrm{ev}},\) and the average error
\(\left\langle\delta \right\rangle_{\mathrm{ev}}=0 .\) The values of \(m\) indicate how efficiently the initial deformation is transformed into the final momentum anisotropy.

In Fig.~\eqref{fig:v2e2} (a) (Top row)  we show the event-by-event distribution of $v_{2}$ vs $\epsilon_{2}$  for $0\%-5\%$ Au+Au collisions at $\sqrt{s_{NN}}=62.4$ GeV. The initial energy density and $\epsilon_{2}$ are obtained from the MC-Glauber model with the contribution coming only from wounded nucleons. The result is obtained for EoS-Lattice, i.e, crossover transition.  Fig.~\eqref{fig:v2e2} (b) (Top row) shows the same, but with EoS having a first order phase transition. Using two different EoS we found $\sim 10\%$ decrease in $c(\epsilon_{2},v_{2})$ for the case of first order phase transition, which clearly indicates that  $c(\epsilon_{2},v_{2})$ can be treated as a good signal of phase transition in the nuclear matter. Above results are not surprising since the speed of sound becomes zero (hence the expansion) for a certain temperature range
 in the first order phase transition. Fig.~\eqref{fig:v2e2} (a,b) (Middle row and bottom row) shows the same results but for $20\%-30\%$ centrality using MC-Glauber and TRENTo model initial conditions. The first order phase transition shows a $\sim 6\%$ and $\sim 5\%$ decrease in the value of $c(\epsilon_{2},v_{2})$ respectively. Thus, we infer that although, the value of $c(\epsilon_{2},v_{2})$ for first-order phase transition is always less than that with crossover transition independent of the model used, the difference is more prominent in the central collisions than at higher centrality. Similarly, other higher order correlations e.g., $c(\epsilon_{n}$ ,$v_{n})$ (for $n=3$ or $n=4$) is found to be smaller for the case of first order phase transition.
 
 However, note that the initial eccentricities $\epsilon_{n}$ are not accessible in real experiments (and are model dependent) and hence the  $c(\epsilon_{n},v_{m})$ are not as interesting as $c(v_{n},v_{m})$ which can be calculated from the available experimental data. As pointed out earlier instead of $c(v_{n},v_{m})$, a clean experimental observable would rather be the normalised symmetric cumulants $\mathrm{NSC}{(m,n)}$.

 This EoS dependence of the flow correlations can be more clearly seen from NSC$(m,n)$ Fig.~\eqref{fig:diffvnvm2030} (left panel), where  we show $\mathrm{NSC}(2,3)$, $\mathrm{NSC}(2,4)$,  and $\mathrm{NSC}(3,4)$ for EoS-Lattice (solid orange circles) and EoS first order phase transition (open blue circles) with corresponding errors for $20\%-30\%$ collision centrality. We  have used the MC-Glauber $\varepsilon_{WN}$ initialisation.  Fig.\ref{fig:diffvnvm2030} (right panel) shows the same but for TRENTo model. The errors are calculated by using bootstrap method.  As can be seen from the Figs.~\eqref{fig:diffvnvm2030} that the $\mathrm{NSC}(2,3)$, $\mathrm{NSC}(2,4)$,  and $\mathrm{NSC}(3,4)$ always distinguishes the two different EoSs. These observations may be attributed to very different evolutionary dynamics of the system for the two different EoS, as the speed  of sound becomes zero in first-order phase transition hence the  linear/non-linear coupling of $\epsilon_{n}$ - $v_{n}$ and $v_{n}$-$v_{m}$ is different in the two scenario. Although the absolute values of the  $\mathrm{NSC}(m,n)$ varies for different initial condition (energy density scales with wounded nucleons or in the TRENTo model), the difference in them remains almost same  for two different EoSs. We found in the mid central collisions $\mathrm{NSC}(m,n)$ is larger for the EoS with first order phase transition irrespective of the initial conditions used here. 

 Finally, in order to see the influence of viscosity in the observables discussed above, we did the simulation for various values of $\eta/s$. In Fig.~\eqref{fig:diffvnvm2030visc}, we show the difference between $\mathrm{NSC}(m,n)$ calculated for EoS- with first order phase transition and that with crossover. The initial energy density is obtained from wounded nucleons $(\varepsilon_{WN})$ in MC-Glauber model for $20\%-30\%$ collision centrality. As in ideal hydrodynamics, the results of $\mathrm{NSC}(m,n)$ for first order phase transition are again larger than that with crossover. Increasing $\eta/s$ increases the difference between them for $\mathrm{NSC}(3,4)$. The same is not true for $\mathrm{NSC}(2,3)$, which is too sensitive to the shear viscosity and responds to it in a rather non-trivial manner.
 
 This indicates that we can utilize  $\mathrm{NSC}(3,4)$ to probe the EoS of the system which implies that one can possibly use this observable to locate the QCD critical point. For example we can calculate  $\mathrm{NSC}(3,4)$ from available experimental data for various $\sqrt{s_{NN}}$ and pinpoint the energies where $c(v_{n},v_{m})$ shows a sudden change in magnitude.
 
\section{\label{sec:Conclusion}Conclusion and outlook}

In this paper, we studied correlation  $\mathrm{NSC}(m,n)$  and  $c(\epsilon_{m},v_{n})$ for $m,n=2-4$ in ultra-relativistic heavy-ion collisions using event-by-event
viscous fluid dynamics model for two different EoSs.  The comparison of numerical results from the hydro code  to the corresponding analytic solution in 1+1D and in 2+1D shows very good agreement between them. Using optical Glauber model as an initial condition, we confirm that our code nicely explains experimentally measured charged pion invariant yield 
and elliptic flow in Au+Au collision at $\sqrt{s_{NN}}=$ 200 GeV and 62.4 GeV. First using ideal hydrodynamics with two different EoS's we showed that the $\mathrm{NSC}(2,3)$, and  $\mathrm{NSC}(3,4)$ clearly distinguishes a first order phase transition scenario and a cross over transition. Although the absolute values of the $\mathrm{NSC}(m,n)$ varies for different initial condition (energy density scales with binary collisions or wounded nucleons) the difference in them  remains almost same  for two different EoSs. We found in the mid central collisions $\mathrm{NSC}(m,n)$ is larger for the EoS with first order phase transition irrespective of the initial conditions used here. In the presence of finite viscosity, the result that the value of  $\mathrm{NSC}(m,n)$ is larger for EoS with first order phase still remains valid. However by inreasing the viscosity  we found the difference between the results from first order phase transition and that with crossover increases for $\mathrm{NSC}(3,4)$.

This indicates that we may utilize $\mathrm{NSC}(3,4)$ to probe the EoS of the system which implies that one can possibly use this observable to locate the QCD critical point. 
For example, we can calculate $\mathrm{NSC}(m,n)$ from available experimental data for various $\sqrt{s_{NN}}$ and pinpoint the energies where $\mathrm{NSC}(m,n)$ shows a sudden
change in magnitude. 

The coefficients $\mathrm{NSC}(m,n)$ can be directly calculated from the available  experimental data, whereas  calculation of $c(\epsilon_{n},v_{m})$ requires input from an initial condition model hence  proves to  be less attractive observable  than $\mathrm{NSC}(m,n)$. However, we would like to point out that our study also shows  $c(\epsilon_{2},v_{2})$ is a promising observable for probing the EoS as it shows $\sim 10\%-30\%$ difference
in the two scenarios. 
\section*{Acknowledgement}
We thank J.Y. Ollitrault and R. Bhalerao for suggestions and helpful discussion.
V.R. is supported by the DST Inspire faculty research grant (IFA-16-PH-167), India. AD acknowledges financial support from DAE, Government of India.
\bibliography{main}

%%%%%%%%%%%%%%% Critical point %%%%%%%%%%%%%%%%%

\end{document}